# Frequency of surveillance testing necessary to reduce transmission of SARS-CoV-2

*Ahmed Elbanna and Nigel Goldenfeld*
*University of Illinois at Urbana-Champaign*
*Urbana, IL 61801*

**Abstract**

We estimate the reduction in transmission of SARS-CoV-2 achievable by surveillance testing of a susceptible population at different frequencies, comparing the cases of both the original Wuhan strain and the Delta variant. We estimate the viral dynamics using viral copy number at first detection combined with considerations arising from aerosol transmission. We take into account the recent findings that infected vaccinated adults may have live viral loads at the same level as infected unvaccinated adults. Our estimates suggest that twice weekly testing, which was adequate for the original strains of SARS-CoV-2 will be insufficient on its own to contain the spread of the Delta variant of concern. We exclude consideration of contact tracing since the rapidity of the onset of viral titre in the case of the Delta variant suggests that unless contact tracing and quarantine are performed very rapidly (ie. much less than a day), these mitigations will be of minimal impact in reducing transmission. These crude estimates do not take into account heterogeneity of susceptibility, social activity, and compliance, nor do they include the additional reduction in transmission that could be achieved by masking and social distancing. In the setting of a large public university, these considerations suggest that risk-targeted testing of vaccinated students, staff and faculty combined with surveillance testing of all unvaccinated individuals is the most efficient way to reduce transmission of COVID-19.

**Overview: the purpose of surveillance testing**

In a recent manuscript, a multi-layered surveillance testing program to mitigate the spread of SARS-CoV-2 in a large public university was outlined [1]. Known as SHIELD, the methodology is effective in reducing cases in a community because it tests everyone on a regular schedule, regardless of whether or not they have symptoms. A positive test result is followed by isolation and quarantine of close contacts. This protocol is known as surveillance testing. The main purpose of surveillance testing is to reduce transmission within a community, by identifying and isolating cases before they have a chance to infect other people. It is not just a diagnostic tool for symptomatic individuals.

In this note, we estimate how frequently it is necessary to surveillance test a population in order to prevent transmission. The results depend on how quickly the virus becomes active in the host, and so we have to treat as separate cases the original Wuhan SARS-CoV-2 virus and subsequent variants. In particular, in this document, we use the viral dynamics to calculate this for the original Wuhan SARS-CoV-2 virus and the Delta variant. From a combination of laboratory data and mathematical analysis we find that twice a week testing is necessary to reduce transmission of the original Wuhan virus, but this is insufficient for the Delta variant. This variant is so transmissible that it requires every other day testing. In fact, it is possible that there will be variants which are so transmissible that surveillance testing on its own is insufficient to reduce transmission. Especially taking into account varying levels of compliance with public health recommendations, a combination of surveillance testing, vaccination, mandatory masking, good ventilation practices and other non-pharmaceutical interventions may be required to control the evolving epidemic.

**Why is surveillance testing successful?**
COVID-19 can of course be transmitted by symptomatic individuals. But it can also be transmitted by individuals who are either asymptomatic (will never present symptoms) or pre-symptomatic (have not yet presented symptoms). Recent analyses on earlier strains of SARS-CoV-2 [2] have demonstrated that transmission by symptomatic individuals represents only 41% of all transmission. The majority of the transmission is from pre-symptomatic individuals (35%) and asymptomatic individuals (24%). Even accounting for uncertainty, the symptomatic transmission is not more than 50%.

The conclusion is that the spread of COVID-19 cannot be achieved by testing only symptomatic individuals. Surveillance testing works because it automatically includes the pre- and asymptomatic individuals, and so has the potential to stop all transmission. In order to realize this potential in practice, we need two requirements:

1. A sensitive test that can detect small amounts of virus before an individual becomes infectious;
2. A frequent test, so that there is a high probability of identifying an infected person before or while they are infectious.

In this brief note, we consider these requirements from two perspectives. First, we look at laboratory data that compares different test modalities and their performance following infection [3]. Second, we look at a mathematical analysis of the reduction of transmission, based upon real-world data on viral dynamics within infected individuals [4-7]. This latter analysis enables us to estimate the effectiveness of surveillance testing on newly-emerging variants, such as Delta, where we do not yet have laboratory data to inform the analysis.

**1. Laboratory data**

**1.1 Test sensitivity**
The sensitivity of the SHIELD saliva test in the two days prior to the start of infectivity ranges from 84% to 88%, much higher than either regular nasal tests (75%) or antigen tests (25-38%) [3]. The SHIELD saliva test will succeed in detecting and subsequently isolating individuals 2 days before they become infectious (and most likely before they show symptoms).

**1.2 Test frequency**
A weekly SHIELD saliva test has a 58% chance of detecting an individual while infectious and a 90% chance of detecting the infection at all [3]. This probability of detecting the individual while infectious rises significantly to between 80-85% if the test is performed twice a week, and the probability of detecting the infection at all rises slightly to 95% [3]!

**1.3 Summary**
For the Wuhan strain of SARS-CoV-2, twice weekly surveillance testing has a significantly higher probability of reducing transmission than testing once a week. Low risk groups could be tested at 1X/week, but 2X/week is more appropriate for riskier groups.

**2. Frequency of testing from measured viral dynamics**

**2.1 Ancestral Wuhan Strain**
We calculated how the basic reproduction number, $R_o$, is modified by a multiplier, *M*, that accounts for the fact that if an individual is detected to be positive and immediately isolated, they are unable to continue infecting others. This results in a fractional reduction of $R_o$ ($R_t = M R_o$) as detailed in Fig. 1 below. Using an infectivity profile that includes pre-symptomatic shedding [4] we could calculate *M* as a function of testing frequency. We found that testing everyone every 7 days yields *M = 0.71* but testing everyone every 3.5 days yields *M = 0.45*, because their infectious period while not isolated (Area A in Fig.1) is reduced.

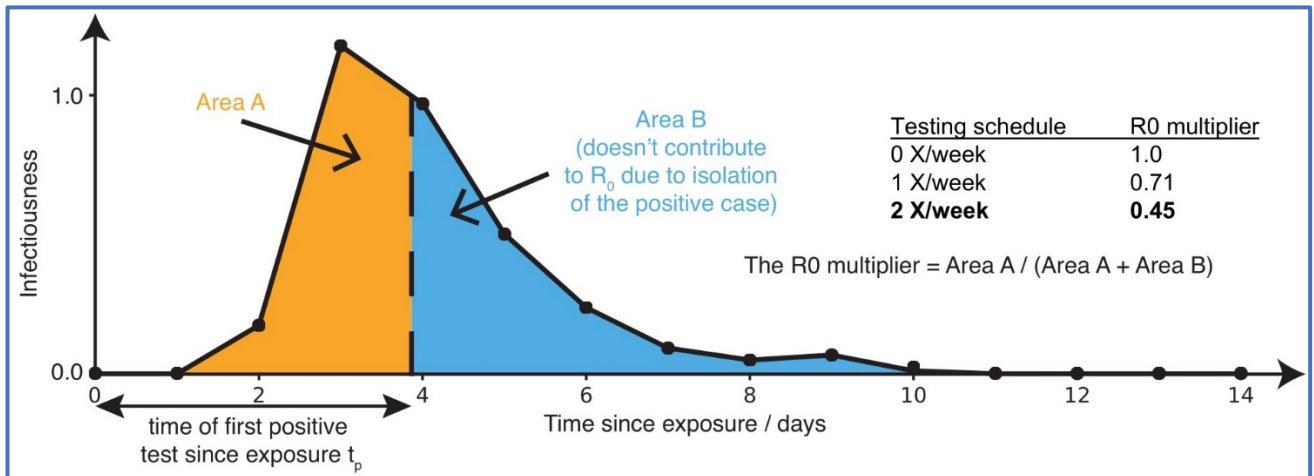

**Fig. 1**: Sensitive testing can reveal a positive case early in the infection, and thus isolation of the index case reduces the number of people infected by this index case. Frequent testing and rapid isolation reduce the time period during which a person is infectious but not isolated (Area A). As a result, the $R_0$ multiplier for testing is the ratio between the truncated area under the curve (Area A) and the untruncated area under the curve (Area A + Area B). The dashed vertical line between Area A and Area B represents the moment an infected individual is isolated; as this line moves to the left, $M$ is decreased, and viral spread is reduced.

## 2.2 The Delta Variant

Unlike in the case of the ancestral strain or the alpha variant, there is only limited information, so far, on the viral dynamics of the emerging Delta Variant. We consider two scenarios for the viral load profile. In the first scenario, we assume that the viral load for Delta has a peaked profile that is qualitatively similar to the viral load of the ancestral strain. In the second scenario, we assume that the viral load for Delta has a plateau at its peak value, as suggested by recent data.

### 2.2.1 Frequency of testing for Delta assuming a peaked viral load profile

In a recent preprint, Li et al [5] investigated the first local transmission of the SARS-CoV-2 Delta variant in mainland China. Using Daily sequential PCR testing of the quarantined subjects, they found that:
- The time from exposure to first detection of the virus (in their daily PCR testing protocol) is about 4 days (IQR 3.00-5.00) in the 2021 Delta epidemic (n=34; peak at 3.71 days) and 6.00 days (IQR 5.00-8.00) during the 2020 epidemic (n=29; peak at 5.61 days) as shown in Fig. 2a
- The viral load in the <u>first positive test</u> for individuals with the Delta variant is about 1000 times larger than the viral load in the <u>first positive test</u> for individuals with the ancestral strain

Together, these findings suggest potentially faster viral replication and greater infectiousness of Delta during the early period of infection. Specifically, according to their analysis of the viral load profile.

1. One day before detection, the viral load may be below detection limit ~ 500 copies/ml.
2. On the day of detection, the viral load may reach about $10^7$ copies/ml.

This growth rate is approximately 4-5 orders of magnitude increase in viral load per day compared to 2-3 orders of magnitude increase in the viral load for the 2020 strain. Furthermore, in the case of the ancestral strain, there exists a period of 1.5-2 days in which while the viral load may go above the limit of detection for a sensitive test before an individual becomes infectious. This window of opportunity is lost in the Delta variant as the individual becomes infectious within less than a day of the onset of rapid viral replication.

Unfortunately, Li et al [5] do not provide information about the peak viral load in the Delta variant cases or the post peak decay rate or the length of the latent and infectious periods. However, more recent work from Chia et al. [6] and Riemersma et al. [7] suggest that the peak viral load for the Delta variant is higher than the ancestral strain and the infectious period may be longer. Furthermore, these studies suggest that the viral dynamics in symptomatic vaccinated individuals is indistinguishable from the viral dynamics in unvaccinated individuals in the first week of illness. These studies, nonetheless, are based on aggregated data and do not provide individualized viral dynamics curves or sufficient information on viral dynamics in asymptomatic vaccinated individuals. Thus, the data currently available on the full viral dynamics of the Delta variant is much more limited than what is available on the other strains.

However, using the available information, we can construct as a working hypothesis, a plausible viral load curve based on:

(i) the initial growth rate provided in Li et al. [5].

(ii) a reasonable decay rate of the viral load in the post peak regime consistent with previous studies on the ancestral strain. For example, a decay rate of one order to one and half order of magnitude drop in the viral load per day has been inferred previously [3] for the ancestral strain. However, the recent work of Chia et al. [6] and Riemersma et al. [7] suggest that the viral load in Delta-infected individuals may plateau near the peak value for at least five days before reportedly declining in vaccinated individuals. Since we are evaluating the worst-case scenario for surveillance testing, we will ignore the observations about the possible existence of viral load plateaus in the calculations that will follow in this section. If confirmed, however, these plateaus will make surveillance testing more effective since the viral load will be detectable for a longer period of time. We will return to this point in the next section.

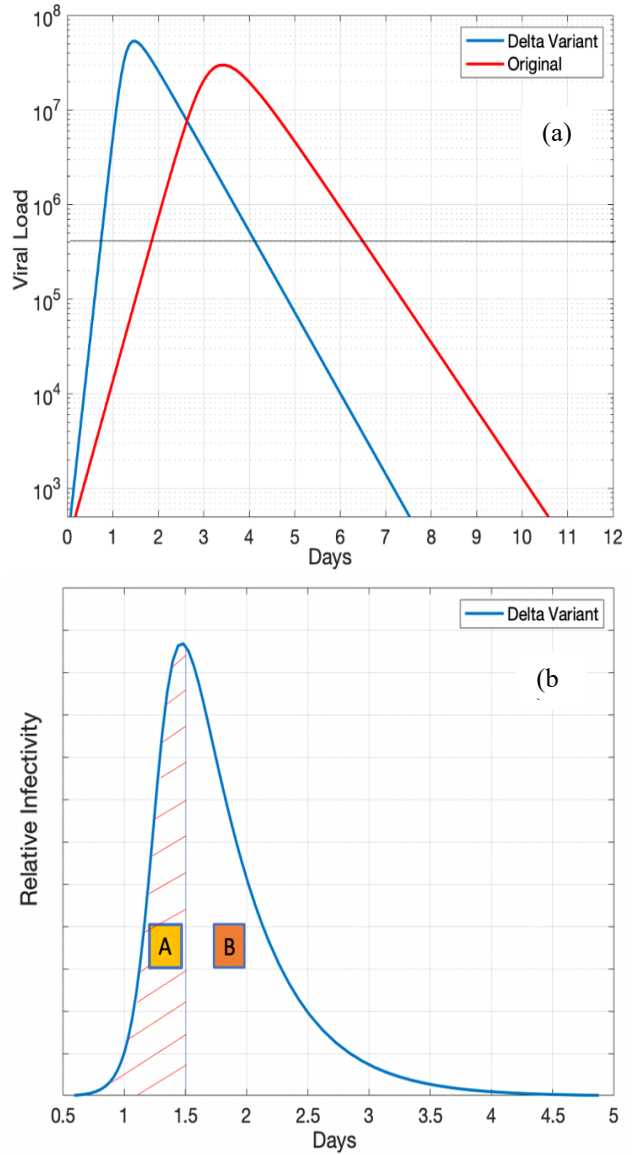

Fig. 2: Schematic for Delta variant characteristics relevant for transmission. (a) A candidate for the viral load dynamics in the case of Delta variant compared to a typical viral load curve of the ancestral strain. (b) Inferred infectiousness profile in the case of Delta variant. The $R_0$ multiplier for testing is the ratio between the truncated area under the curve (Area A) and the untruncated area under the curve (Area A + Area B). The dashed vertical line between Area A and Area B represents the moment an infected individual is isolated; as this line moves to the left, $M$ is decreased, and viral spread is reduced.

(iii) the observation that the inferred $R_o$ for Delta is about as twice as large as that of the ancestral strain. A candidate for a probable viral load profile in the case of Delta variant is shown in Fig. 3a below, compared with a typical viral load profile for the original strain;
(iv) our knowledge that COVID-19 is airborne and spread by aerosols. Buonanno et al. [8] provided empirical scaling formulae that connects the viral load with the infectious quanta that may be emitted by an infected individual during breathing or speaking.

Guided by these observations, we transform the viral load into infectivity and plot an example of the infectiousness profile for the Delta variant in Fig. 3b.

Using the same methodology outlined above for the ancestral strain, we compute how the basic reproduction number $R_o$ for the Delta variant is modified by a multiplier $M$ due to frequent testing and immediate isolation upon detection. We found that: (i) testing every 7 days yields $M = 0.8$, (ii) testing every 3.5 days yields $M = 0.68$, but (ii testing every 2 days yields $M = 0.36$. The results are summarized in Table 1.

**2.2.2 Frequency of testing for Delta assuming a plateau-like viral load profile**

Kang et al. [9] estimated the epidemiological characteristics of Delta using transmission dynamics from an outbreak in Guangdong, China, in May-June 2021. The mean estimates of the latent period and the incubation period in their studies were 4.0 days and 5.8 days, respectively. Consistent with earlier work by Li et al. they also found evidence for rapid initial viral replication. Furthermore, they estimated that about 74% of the infections occur during the pre-symptomatic period and before symptom onset. The viral load also appears to plateau near the peak for up to six days after symptom onset before it starts to gradually decrease. Using all this data, we may construct a relative infectiousness profile as shown in Fig. 3. Here, an individual becomes infectious between 3 and 4 days after exposure, quickly reaching peak infectivity consistent with the rapid viral growth rate characteristic of delta, then remains at that peak for several days. We assume that symptoms appear at day 6 after exposure. To satisfy that about 74% of the transmission occurs in the pre-symptomatic period, we assume that the infected individual self-isolate within a day following the onset of symptoms. If that does not happen, an increased proportion of symptomatic transmission will occur. Note that since the viral load in infected individuals has a plateau-like profile and does not decay rapidly after symptom onset as in the ancestral strain, the basic reproduction number of the Delta variant might vary significantly depending on the date of isolation. This may explain the large uncertainty in estimating R0 with upper values exceeding 9 in some cases.

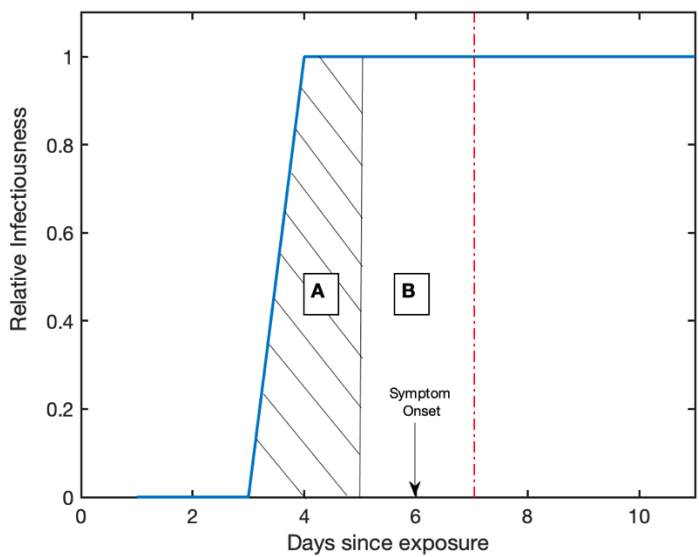

Fig. 3: Inferred infectiousness profile for Delta following epidemiological parameters in Kang et al. [9].

Using a similar approach as outlined before, we compute how the basic reproduction number $R_o$ for the Delta variant is modified by a multiplier $M$ due to frequent testing and immediate isolation upon detection. We found that: (i) testing every 7 days yields $M = 0.75$, (ii) testing every 3.5 days yields M

= 0.44, but (iii) testing every 2 days yields M = 0.28. These calculations are conditioned on the assumption that, in the absence of frequent testing, an individual will most likely self-isolate within a day after onset of symptoms. If there is no self-isolation, the individual will continue to transmit the virus with peak infectivity. The effect of this will be that frequent testing will be more efficient as the multiplier will be reduced, since Area B in Figure 3 will increase. The downside, however, is that this multiplier will be used with a higher basic reproduction number since delays in isolation will result in infecting more people.

Table 1: Effect of testing frequency on the multiplier for the basic reproduction number

| Testing Frequency | Ancestral Strain | The Delta Variant (Peaked Profile) | The Delta Variant (Plateau Profile) |
|---|---|---|---|
| Every 7 days | 0.70 | 0.80 | 0.75 |
| Every 3.5 days | 0.45 | 0.68 | 0.44 |
| Every 2 days | 0.30 | 0.36 | 0.28 |

These findings, despite limited available data and multiple sources of uncertainty, highlight the challenges in controlling Delta variant transmission irrespective of whether the viral load has a single peak or a plateau. Specifically:
- Higher testing frequency, compared to the case of the ancestral strain, is needed. The minimum frequency for the Delta variant is testing every other day to cope with the fast viral replication and increased infectivity during early infection.
- Speed in returning test results and isolation of infected individuals cannot be over emphasized enough in the case of Delta variant. Delays in isolation or lack of compliance with public health measures are potentially very damaging to efforts of controlling transmission due to the rapid viral dynamics and higher viral loads.

**Summary**

In essence, what these calculations reveal is that to attain the same level of mitigation that SHIELD provided for the ancestral Wuhan strain, keeping all other mitigations the same, one needs to test every other day for the Delta variant, whereas twice a week was adequate for the Wuhan strain.

**Discussion**

The evolution of SARS-CoV-2 virulence is certainly affected by the selection pressure to which it is subject. During the emergence of the Wuhan strain, there was no vaccine, and the selection pressure would have come primarily from non- pharmaceutical interventions. Since the virus is transmitted during the asymptomatic or pre-symptomatic stage, there is no reason to expect it to become more temperate, as is usually the case for some other viruses. The main variant of concern that emerged was B117 or Alpha, which was about 50% more transmissible but apparently not more virulent.
During the emergence and spread of Delta, the vaccination campaign was going on. Vaccines elicit an immunizing response that will cause a rapid diminution of viral titre upon infection. Thus, one would expect variants to be selected which create large amounts of viral titre before the immune response kicks in. We speculate that this might be what have happened with Delta. The outcome is that vaccinated and unvaccinated people can transmit the virus, although the vaccinated people will have a shorter window of transmission.

**Conclusion**

For the Wuhan strain of SARS-CoV-2, twice weekly surveillance testing has a significantly higher probability of reducing transmission than testing once a week. Low risk groups could be tested at 1X/week, but 2X/week is more appropriate for riskier groups.

For the Delta variant, testing unvaccinated individuals every other day has a significantly higher probability of reducing transmission than testing once or twice a week. The higher basic reproduction number of the Delta variant can be estimated from a variety of early reports and is summarized in a meta-analysis [10]. The results for $R_o$ can range as high as 8 with a plausible value in the range 5-6. This then suggests that testing every other day as a sole mitigation strategy will not be sufficient to control transmission because the effective reproduction number using surveillance testing on its own will yield $MR_o$ = 0.36 x 6 = 2.16, for the peaked viral load profile, and will yield $MR_o$ = 0.28 x 6 = 1.68, for the viral load profile with a plateau, in other words well above 1. Thus, surveillance testing of a susceptible population on its own will not prevent transmission. A layered mitigation approach including expanded vaccination coverage, universal KN95 or KN94 masking, and improved ventilation should be implemented.

Since the viral dynamics in symptomatic vaccinated and unvaccinated individuals appear to be similar at least in the first six days after admission to hospital [6], targeted testing for vaccinated individuals who reside in locations with high prevalence combined with surveillance testing for unvaccinated individuals may be an effective strategy for reducing transmission. The reason is that vaccinated individuals still have lower probability than unvaccinated individuals of getting infected in the first place, but the precise quantification of this risk reduction is still unknown. Furthermore, infectiousness of asymptomatic vaccinated individuals is not established yet, although asymptomatic and pre-symptomatic unvaccinated individuals transmit the virus. Population-wide surveillance testing of vaccinated individuals will be undesirable and would certainly be faced with lower compliance and may even deter the highest public health priority of expanding vaccination coverage. Therefore, targeted testing of vaccinated individuals who are most exposed to getting infected through their residence may be more appropriate than surveillance testing, be more cost-effective, and would effectively complement a multilayer mitigation approach including mask mandate, proper ventilation, and social distancing.